\documentclass[conference]{IEEEtran}
\IEEEoverridecommandlockouts
\usepackage{pgfplots}
\usepackage{stfloats}
\usepackage{colortbl}
\usepackage{multirow}
\usepackage{hhline}
\usepackage{cite}
\usepackage{amsmath,amssymb,amsfonts}
\usepackage{algorithmic}
\usepackage{graphicx}
\usepackage{textcomp}
\usepackage{xcolor}
\usepackage{stackengine}
\usepackage{float}
\usepackage{subfig}
\usepackage{multicol}
\usepackage{hyperref}
\usepackage[inkscapeformat=png]{svg}
\usepackage[a4paper, total={184mm,239mm}]
{geometry}
\def\BibTeX{{\rm B\kern-.05em{\sc i\kern-.025em b}\kern-.08emT\kern-.1667em\lower.7ex\hbox{E}\kern-.125emX}}

\begin{document}

\newcommand{\mv}[1]{{\color{black}#1}}
\newcommand{\sr}[1]{{\color{yellow}#1}}
\newcommand{\jc}[1]{{\color{blue}#1}}

\newcommand{\red}{\textcolor{black}}
\newcommand{\blue}{\textcolor{blue}}
\newcommand{\Ttline}{\rowcolor{red!8}}
\newcommand{\Tgrline}{\rowcolor{darkgray!10}}
\newcommand{\Tggline}{\rowcolor{green!15}}
\newcommand{\Tgglinee}{\rowcolor{green!8}}
\newcommand{\Tbline}{\rowcolor{blue!15}}
\newcommand{\redt}{\rowcolor{red!8}}
\newcommand{\greent}{\rowcolor{green!8}}
\newcommand{\bluet}{\rowcolor{blue!8}}
\newcommand{\V}{$\scriptsize\checkmark$}
\newcommand{\X}{$\scriptsize -$}
\newcommand{\mr}{\multirow}
\newcommand{\mc}{\multicolumn}
\newcommand{\tbf}{\textbf}
\newcommand{\TM}{\textsuperscript{\tiny{TM}}}

\title{\vspace{-0.2cm}TreeGRNG: Binary Tree Gaussian Random Number Generator for Efficient Probabilistic AI Hardware\vspace{-0.2cm}}

\author{\IEEEauthorblockN{Jonas Crols, Guilherme Paim, Shirui Zhao, Marian Verhelst}
\IEEEauthorblockA{\textit{MICAS-ESAT, KU Leuven, Belgium}}}

\maketitle

\begin{abstract}
Bayesian Neural Networks (BNNs) offer opportunities for greatly enhancing the trustworthiness of conventional neural networks by monitoring the uncertainties in decision-making.
A significant drawback for BNN inference at the extreme edge, however, is the imperative need to incorporate Gaussian Random Number Generators (GRNG) within each neuron. 
State-of-the-art GRNG algorithms heavily depend on multiple arithmetic operations and the use of extensive look-up tables, posing significant implementation challenges for ultra-low power hardware implementations.
To overcome this, this paper presents an innovative binary tree random number generator (TreeGRNG) allowing the use of ultra-low-cost constant comparators instead of arithmetic units.
We further enhance the TreeGRNG proposal with a set of hardware-aware optimizations exploiting the Gaussian properties.
The optimized TreeGRNG surpasses the State-of-the-Art (SoTA) in terms of distribution accuracy while achieving a 3.7$\times$ reduction in energy per sample and boosting the throughput per unit area by 5.8$\times$.
Moreover, our TreeGRNG proposal possesses a distinct advantage over the current SoTA in terms of flexibility, as it easily enables designers to adjust the shape of the sampled probability distribution, extending beyond the capabilities of traditional GRNGs, opening the horizon towards future probabilistic AI designs. The TreeGRNG design is available open-source in the link\footnote{\url{https://github.com/KULeuven-MICAS/TreeGRNG}}.
\end{abstract}


\section{Introduction}\label{intro}

Throughout the years, Random~Number~Generators~(RNGs) have consistently demonstrated their utility in a variety of applications, including probabilistic inference, Monte-Carlo simulations\cite{b1}, and financial modeling.  They excel in situations where introducing a controlled level of unpredictability is essential to produce resilient and reliable results.

Recently, Gaussian~RNGs~(GRNGs) have gained significant importance in the realm of Bayesian~Neural~Networks~(BNNs). BNNs expand upon the capabilities of conventional neural networks by incorporating the capacity to quantify uncertainties.
The uncertainty quantification provides the BNNs the immense potential to revolutionize safety-critical domains, such as healthcare assistance and autonomous~driving~\cite{b12}.

Executing BNN inference at the extreme edge, however, presents a notable obstacle: the essential requirement of integrating GRNGs into each individual neuron to introduce uncertainty on the weights comes with significant area and energy overheads, conflicting with the resource scarcity of extreme edge platforms.

Recent works in this field have explored several algorithms for GRNGs, such as the Table-Hadamard \cite{b1}, 
Ziggurat~\cite{b2}, and the Box-Muller~\cite{b3} algorithms. 
Unfortunately, these GRNG algorithms, all still demand complex functions and a large number of arithmetic operations.
Consequently, their throughput remains suboptimal, and their hardware costs are prohibitively high, limiting their practicality in the context of BNNs.
 
\begin{figure}[!t]
    \centering
    \includegraphics[width=.92\linewidth]{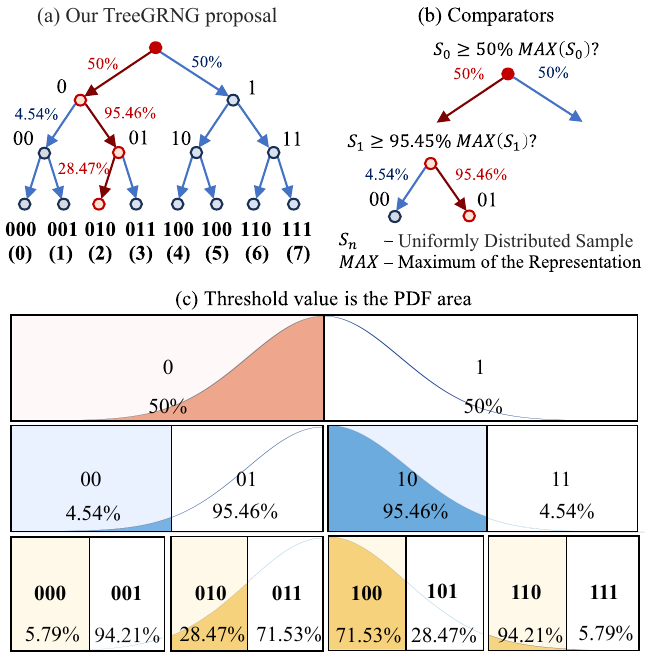}
    \vspace{-0.15cm}
    \caption{\small{(a) Our TreeGRNG proposal, the (b) comparators, and (c) the threshold calculation based on the PDF curve area.}} 
    \vspace{-0.6cm}
    \label{fig1}
\end{figure}

To address this challenge, this paper presents the Binary~Tree~GRNG~(TreeGRNG), as depicted in Fig.~\ref{fig1}-a. 
Our algorithm substantially simplifies the GRNG hardware implementation by using low-cost constant threshold comparators. 
This brings a key advantage to our TreeGRNG proposal, as it reduces the overall circuit area by around 10$\times$ in contrast to the State-of-the-Art (SoTA) designs in which the algorithms rely on complex functions, arithmetic units, and look-up tables.

Fig.~\ref{fig1}-a showcases the binary tree representation of our proposal.
Within our TreeGRNG proposal, multiple sorting experiments occur in a cascade, with each event probability influenced by all preceding events.
Through this method, we can acquire any discrete distribution by tailoring the probabilities for each split. 
The thresholds for each decision node are pre-computed using the integral of the probability density function (PDF) curve over equidistant intervals that each take $1/2^{L+1}$ of the range (Fig.~\ref{fig1}-c), where L is the tree level.
Our TreeGRNG algorithm allows a highly optimized hardware design that outperforms the SoTA in distribution accuracy, delivering an impressive 3.7$\times$ reduction in energy and a substantial 5.8$\times$ increase in throughput per unit area.

Section~\ref{rw} reviews the related work, while 
Section~\ref{algorithm} introduces our TreeGRNG algorithm. 
An efficient TreeGRNG hardware implementation is shown in Section~\ref{architecture}, followed by a set of optimization techniques that exploit useful characteristics of the Gaussian distribution to further simplify the hardware.
Finally, Section~\ref{results} compares the results to the SoTA.

\section{Related work}\label{rw}

As previously mentioned, GRNGs have been implemented using various algorithms, with some of the most popular ones being the Table-Hadamard, the Ziggurat, and the Box-Muller.

The Ziggurat operates as a rejection algorithm, randomly generating a point within a plane that encompasses the target distribution. 
It subsequently examines whether this point falls within the desired distribution and retries if it is out.
The work presented in \cite{b2} demonstrated a Ziggurat hardware implementation. 
To mitigate complexity, a piecewise linear approximation is employed in \cite{b2}, yet its performance remains constrained due to the reliance on many arithmetic operations, such as adders and multipliers.

The Box-Muller algorithm is a method that represents an improvement over the inverse transform technique. 
The idea behind Box-Muller is to transform a pair of Uniformly Random Numbers (URNs) into two independent Gaussian Random Numbers (GRNs). 
The algorithm requires the computation of complex arithmetic functions for a given sample, like the natural logarithm, sine, cosine, and square root.
Puntsri's work, as detailed in \cite{b3}, demonstrates the hardware implementation of this algorithm by employing polynomial approximations for the natural logarithm, Lookup Tables (LUTs) for sine and cosine calculations, and a CORDIC-based algorithm for square root computation.
Each of these components introduces complexity and cost negatively impacting its overall efficiency.

Recently, a new architecture was designed by Dorrance\cite{b6}. 
This architecture uses the table-Hadamart Transform (HT) \cite{b6} and improves it by time-interleaving (TI) each step. 
This TI-HT architecture \cite{b6} has the advantage of reducing arithmetic requirements in terms of additions and subtractions. However, this reduction costs additional memory \mv{accesses, as well as requiring} a multi-cycle architecture.

\textbf{Distinction from SoTA:} 
Table~\ref{tab:RelatedWorkResults} shows a summary of each related work characteristic with a comparison with our TreeGRNG proposal.
Our proposal brings a novel approach based on binary trees, eliminating the necessity for intricate functions, arithmetic units, lookup tables, or extra memories. 
TreeGRNG is a versatile solution with universal applicability, capable of approximating diverse PDF shapes that go beyond the Gaussian distribution.
Its hardware implementation also allows a fine-tuning of the tradeoff between cost and distribution accuracy to align with the application requirements.

\begin{table}[t]
\vspace{-0.1cm}
\centering
\caption{\small{Summary of the related work.}}
\label{tab:RelatedWorkResults}
\vspace{-0.1cm}
\setlength\extrarowheight{.4pt}
\begin{tabular}{| c || c | c | c | c | c | c | c | }
\hhline{~-------}
\Ttline \multicolumn{1}{c|}{\ }  & \multicolumn{7}{|c|}{\textbf{Summary Results}} \\ 
\hline\hline
\Ttline Related Work                    &  A   &  B  &  C  &  D &  E & F & G  \\ \hline
         Ziggurat   \cite{b2}           & \V   & \X  & \X  & \X & \V & \X & \X   \\
\Tgrline Box-Muller \cite{b3}           & \X   & \X  & \X  & \V & \X & \X & \X   \\
         TI-HT      \cite{b6}           & \V   & \V  & \X  & \X & \X & \X & \X   \\
\Tggline   \tbf{Our TreeGRNG}           & \V   & \V  & \V  & \V & \V & \V & \V   \\  
\hline\hline
\end{tabular}
\vspace{0.15cm}
\setlength\extrarowheight{.2pt}
\begin{tabular}{cccccccc}
\mc{8}{l}{ \ \scriptsize (A) Avoids complex functions. (B) Avoids multipliers.} \\
\mc{8}{l}{ \ \scriptsize (C) Avoids adders. (D) Avoids Look Up Tables.}  \\
\mc{8}{l}{ \ \scriptsize (E) Memory-less implementation.}  \\
\mc{8}{l}{ \ \scriptsize (F) Universally applicable for any PDF distribution.} \\
\mc{8}{l}{ \ \scriptsize (G) Extra optimizations for symmetric PDF distributions.} \\
\end{tabular}
\vspace{-.7cm}
\end{table}

\section{The Binary Tree GRNG Algorithm Proposal}\label{algorithm}
Our GRNG algorithm aims to generate integer Gaussian Random Numbers with a design-time customizable N-bit precision.
We first show the proposed TreeGRNG algorithm to approximate any PDF distribution. Afterwards, we  show how to use this technique to produce the Gaussian distribution.\\
The proposed algorithm consists of two steps: 1.) bin generation, which is done once at design time; and 2.) bin selection, which happens for each sample at run-time, driven by uniform random bits.

\subsection{Bin creation}\label{bins}
Inspired by binary trees, our TreeGRNG algorithm splits the numeric output range into equally-sized bins, each represented by a binary index number, as visualized in Fig.~\ref{fig1}-c. The total range gets split by two, and subsequently, each section gets split again. This process is repeated until the desired amount of bins (dictated by the desired output resolution) is reached.

As visualized in the red-colored path in the tree of Fig.~\ref{fig1}-a, the output indices in bold at the leaf node can be obtained by a unique path taken through the binary tree to reach a certain bin. 
This can now be exploited to ensure that each output value at the bottom of the tree is reached with a probability corresponding with the desired PDF. 

To this end, each subsequent level $i$ of the binary tree splits the output range into two. In each binary tree node, a decision to the left/right side of the node takes place and adds a logic '0'/'1' to the index, respectively. 
(see the nodes in Fig.~\ref{fig1}-a).
The probability of moving left or right at a stage $i$ should be influenced by the shape of the PDF, as depicted in Fig.~\ref{fig1}-c for the normal distribution. Specifically, the required probability in each binary tree node to go left/right corresponds to the area under the PDF of the left, resp. right bin. 
For an N-bit indexing scheme, this process gets repeated N-1 times, each level adding one extra bit to the final output sample.
Using this technique, the probability of ending up in any lowest-level bin is determined by the chances of going left or right in every binary tree node, while the bin binary index can be easily derived from the path taken to reach the bin. 
The number of levels necessary to form all paths is equal to the number of bits \mv{desired} for the output sample \mv{to achieve the target resolution}. 
Any PDF distribution with a total of $2^{N}$ bins can be represented with $N$ levels.

\subsection{Bin selection}\label{selection}

When traversing the tree at run-time, a weighted coin flip is required in each binary tree node to pass to the left or the right of the node, until a leaf is reached and an output index is obtained. 
However, the chance that a certain bin gets randomly selected should be equal to the bin's respective probability on the desired PDF curve. 
This is achieved by manipulating the binary tree nodes such that they are either more or less likely to send samples to the left or right. 

A uniform random number generator (URNG) is used to \mv{realize this} behavior.
Uniform random numbers are generated at run-time and compared with a threshold to 
determine whether to go left or right.
The decision threshold for each node is \mv{different and} determined at design time using the area under the target PDF curve (i.e., the integral of the PDF in the bin's limits), see Fig.~\ref{fig1}-c.
The probability $P$ in (1) of going left in stage $i$ is determined as the ratio of the area under the curve of the left bin of stage $i+1$ in Fig.~\ref{fig1}-c, compared to the encompassing bin of stage $i$: 

\begin{equation}
    \label{probability}
\small   P(left~split) = \frac{Area_{left}}{Area_{combined}}
\end{equation}

In Fig.~\ref{fig1}-a the probabilities for the Gaussian are shown as an example for each split made along a random path. 
The result after $N$ tree levels will be $2^N$ bins outputting Gaussian samples with mean $2^{N-1}$ and sigma $2^{N-3}$.
Each level that is added improves the resolution 
of the generated distribution.\\ 

\subsection{Gaussian PDFs}\label{gauss}
To employ our new algorithm for generating GRNs, the Gaussian PDF curve is used to determine the thresholds. 
As its range is normally infinitely large, a maximum and minimum value needs to be chosen so that it can be split into equal-sized bins. 
Larger maximal values provide longer tail lengths, which is important in applications such as Monte Carlo simulations. 
Improvements in tail length go along with compromises in accuracy in the main blob as only a finite set of $2^N$ bins can be represented. 
A good balance can be found depending on the application where both tail length and the distribution accuracy in the main blob are satisfactory. 
All tests of this paper are executed with an implementation that uses a maximum tail length of 4$\sigma$ so that an 8-bit random sample will use the fixed$\langle$3,5$\rangle$ 2's Complement notation.
To generate a configurable $\mu$ and $\sigma$, an adder and multiplier unit can be deployed to transform its output, as for any GRNG of the related work.

\section{TreeGRNG Hardware Design}\label{architecture}
This section discusses the hardware-efficient implementation of the TreeGRNG algorithm. 
We will first describe the general architecture that embraces any PDF, then, we further detail a set of hardware optimizations exploiting the specific characteristics of a Gaussian PDF.

\subsection{Binary tree implementation}\label{concept}
In order to implement the TreeGRNG algorithm, two basic blocks are used (a) a URNG and (b) a set of constant comparator units that uses the URNs to emulate a weighted coinflip. We explain both blocks in the following subsections.

\subsubsection{URNG}
The proposed TreeGRNG architecture uses Linear Feedback Shift Registers (LFSR) \cite{b4} to provide pseudo-random numbers. 
Integrating LFSRs aligns well with our resource-constrained design requirements, as they can deliver high-quality URNs at elevated throughput rates while maintaining the hardware cost limited.
LFSRs can be modeled as Galois fields, meaning that they return a seemingly random sequence of uniformly distributed numbers.
In reality, this sequence is finite and repeats itself. 
A K-bit LFSR will repeat itself after $2^K-1$ iterations. 
Given that every level employs an LFSR, high values for $K$ will be expensive to implement in the architecture. 
However, if the order of the GRNG is too small, its quality will significantly fall.
To face this conflict, our GRNG uses different length LFSRs at each level characterized by a large least common multiple (lcm), to increase the total order of randomness.
Two LFSRs of length $K$ and $L$ have a combined order growing rapidly:

\begin{equation}
    \label{lfsrord}
 \small   ord(GRNG) = lcm(2^K-1,2^L-1)
\end{equation}

Note that it is also important that no two LFSRs of the same length can have the same seed value, as this would break the randomness among the binary tree nodes and result in gaps and spikes in the final distribution. After a reset has occurred, a single clock cycle is needed to reload this unique seed value.

\subsubsection{Weighted Coinflip}

The weighted coin flip in each binary tree node to go left or right is implemented with comparators. 
In each node, the uniform random number generated by an LFSR is compared to a pre-determined threshold, generating a logic '1' or logic '0' output bit as discussed in Section~\ref{selection}.
This quantized threshold gets calculated by multiplying the probability from equation (1) 
with the largest number of the representation from the LFSR output with length $K$. For a $K$-bit URN, the quantized threshold will be equal to: 

\begin{equation}
    \label{threshold}
 \small   T = \frac{Area_{left}}{Area_{combined}}\cdot(2^K-1)
\end{equation}

For \mv{a known targeted} distribution, these predetermined thresholds \mv{for tree node} can be constant. 
These constant thresholds \mv{can therefore be determined} once at design time, and \mv{hardcoded}. \mv{As a result,} the generic comparator \mv{in each node of the tree can be} replaced by ultra-low-cost constant comparators. 
This allows the synthesis tool to efficiently simplify the comparator block to what is necessary, cutting down on hardware cost significantly\cite{b5}. 

The \mv{use of such} efficient constant comparators \mv{is possible for the generation of any distribution which is fully known at design time.} GRNGs with configurable $\mu \neq 0$ and $\sigma \neq 1$ \mv{can be built out of a fixed uniform GRNG, followed by an offset and gain adjustment stage}, as discussed in Section~\ref{gauss}. 

\begin{figure*}[htbp]
    \centering
    \vspace{-1.6cm}
    \includegraphics[width=.83\textwidth]{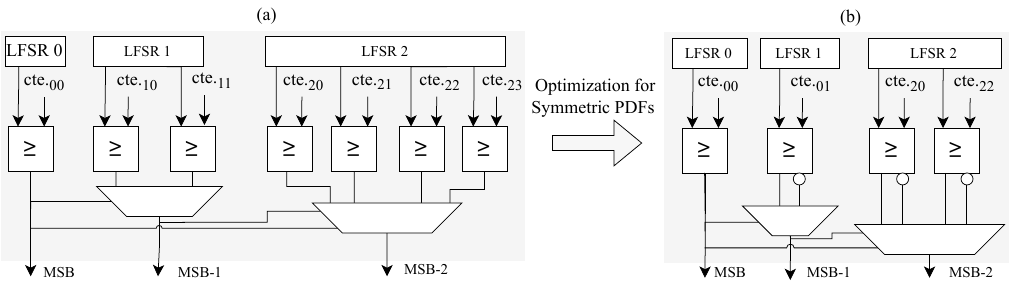}
    \caption{\small{The three-level example for the (a) general TreeGRNG architecture and its (b) optimized version for symmetric PDFs.}}
    \label{archit}
    \vspace{-.4cm}
\end{figure*}

\subsubsection{Integrating all the components}

The different LFSRs and decision blocks of each tree level are integrated in a parallel \mv{implementation,} to efficiently realize the complete binary tree. 
The complete TreeGRNG architecture is visualized in Fig.~\ref{archit}-a, and is generic for any \mv{targeted} PDF distribution. 
It consists of one LFSR stage per level $i$, together with \mv{$2^i$} decision blocks \mv{per level}.
Within each level, \mv{each decision node requires a different threshold for its comparator}.
Hence, a multiplexer is required to select the correct threshold based on the \mv{decision} results of the previous levels. 
This multiplexing stage can be positioned before or after the comparators, with the former resulting in a slightly more area-efficient solution, and the latter approach in an improved throughput \mv{(i.e., shorter critical path)}. \\ 
As shown in Fig.~\ref*{archit}, the resulting architecture allows the generation of a new GRN \mv{sample} in a single clock cycle. 

\subsection{Further \mv{GRNG-specific} hardware efficiency optimizations}\label{improv}
While the implementation discussed in the previous subsection is capable of generating samples from any discrete distribution with $N$ bits, further efficiency optimization can be exploited when targeting \mv{a Gaussian PDF specifically,} 
to cut down on hardware cost while minimizing loss in accuracy.
 
In the baseline implementation, the number of comparators required to represent all possible thresholds grows exponentially with the tree level. 
Luckily the Gaussian PDF provides excellent characteristics that can 
reduce the hardware cost and make the algorithm suitable for large bitwidth GRN. 

\textbf{1.) Symmetry optimization:} First, the designer can exploit the fact that a uniform Gaussian has symmetry \mv{along} the y-axis.
This means that bins \mv{on the left and on the right side of the curve can reuse the same thresholds. As a result, it is possible to} 
replace the comparators for bins on the right side of the curve with the inverted output of the corresponding bin on the left side.
This easy simplification reduces the number of comparators by a factor of two, without suffering any compromise in accuracy. 
An architecture that takes advantage of this optimization is shown in Fig.~\ref{archit}-b.

\textbf{2.) Cluster optimization:} As a second optimization, \mv{it can be exploited that} the thresholds found in neighboring bins tend to be \mv{similar (or sometimes identical)} for large values of $i$. \mv{This allows to further reuse the comparators across different nodes of the same level, to the benefit of a lower hardware cost. There is of course a trade-off between the number of clustered thresholds and the accuracy of the generated distribution.}

\begin{table}[htbp]
    \centering
\setlength\arrayrulewidth{0.1pt}
    \setlength\extrarowheight{.3pt}
    \caption{\small{Symmetry and Cluster Optimization Benefits}}
    \vspace{-0.1cm}
    \begin{tabular}{ c c | c c c|}
    \hhline{~~---}
    & & \textbf{Baseline} & \textbf{Symmetry} & \textbf{Symmetry + } \\
    & & \textbf{TreeGRNG} & \textbf{Opt.} & \textbf{Cluster Opt.} \\
    \textbf{Level}&\textbf{Cluster size}&\multicolumn{3}{c|}{\textbf{Number of Comparators}} \\
    \hline
    0&1&1 & 1&  1 \\
   \Tgrline  1&1&2 & 1& 1 \\
    2&1&4 & 2 & 2 \\
  \Tgrline   3&1&8 &  4&  4 \\
    4&1&16 & 8& 8 \\
   \Tgrline  5&1&32& 16&  16 \\
    6&2&64 & 32& 16\\
   \Tgrline  7&8&128& 64 & 8 \\
    \hline\hline
    \multicolumn{2}{r|}{\textbf{Total Count}} & 255& \multicolumn{1}{c}{128}& \multicolumn{1}{c|}{56} \\
    \Tgglinee \multicolumn{2}{r|}{\textbf{Reduction}} & - & \multicolumn{1}{c}{\textbf{$-1.99\times$}}& \multicolumn{1}{c|}{\textbf{$-4.55\times$}} \\
    \hhline{~----}
    \end{tabular}
    \label{compam}
    \vspace{0.2cm}
    \begin{tabular}{llll}
    \mc{4}{l}{ \ \ \ \ \small *This example is designed to generate 8-bit samples.} \\
    \end{tabular}
    
    \vspace{-0.1cm}
\end{table}

Table~\ref{compam} shows the proposed sizes of the threshold clustering \mv{(number of adjacent nodes which make use of the same threshold and hence same comparator)} for an architecture yielding an 8-bit result.
The amount of grouping possible depends on both the difference in thresholds between neighboring bins and the \mv{desired distribution} accuracy. 
Deeper tree levels have their thresholds closer to each other, so more grouping can be performed with minimal error. We will \mv{show in Section~\ref{results}-A} that \mv{the proposed clustering of Table II results in} only very minor approximation errors. 

\mv{As can be seen in Table II, for the proposed 8-bit GRNG clustering,} Level 6 is able to reduce the amount of thresholds from 64 to 16, while level 7 can go even further and requires only 8 total thresholds instead of 128. 
This is possible considering thresholds at these levels are so similar they correspond to the same number after quantization.
Note that this simplification does not mean that neighboring bins have an equal likelihood of appearing. 
This means that paths of neighboring bins have equal probabilities of going in a certain direction. 
This optimization transforms the total complexity of the architecture from growing exponentially to growing only linearly.
The table also indicates the final number of comparators per level from exploiting symmetry as well as identical decision thresholds.
The used cluster sizes \mv{can be adapted upon the desired trade-off between hardware cost and generated distribution accuracy.} 

\textbf{3.) Pipelining:} Lastly, given that the datapath is fully combinatorial except for the LFSRs, the datapath can be pipelined \mv{to increase throughput}. 
In Fig.~\ref*{archit}, the critical path starts with LFSR 0 and goes through one mux per level, to finally end at the LSB of the result.
Pipeline registers can be placed behind the comparators to portion off the comparison stage from the selection stage.
For long bitwidths, additional registers can be added in the middle of the multiplexer selection stage to further split the critical path.

\section{Experimental results}\label{results}

\subsection{Algorithmic error evaluation}\label{error-eval}

\begin{figure*}[!t]
    \vspace{-1.6cm}
    \centering
    \centerline{\includegraphics[width=.83\textwidth]{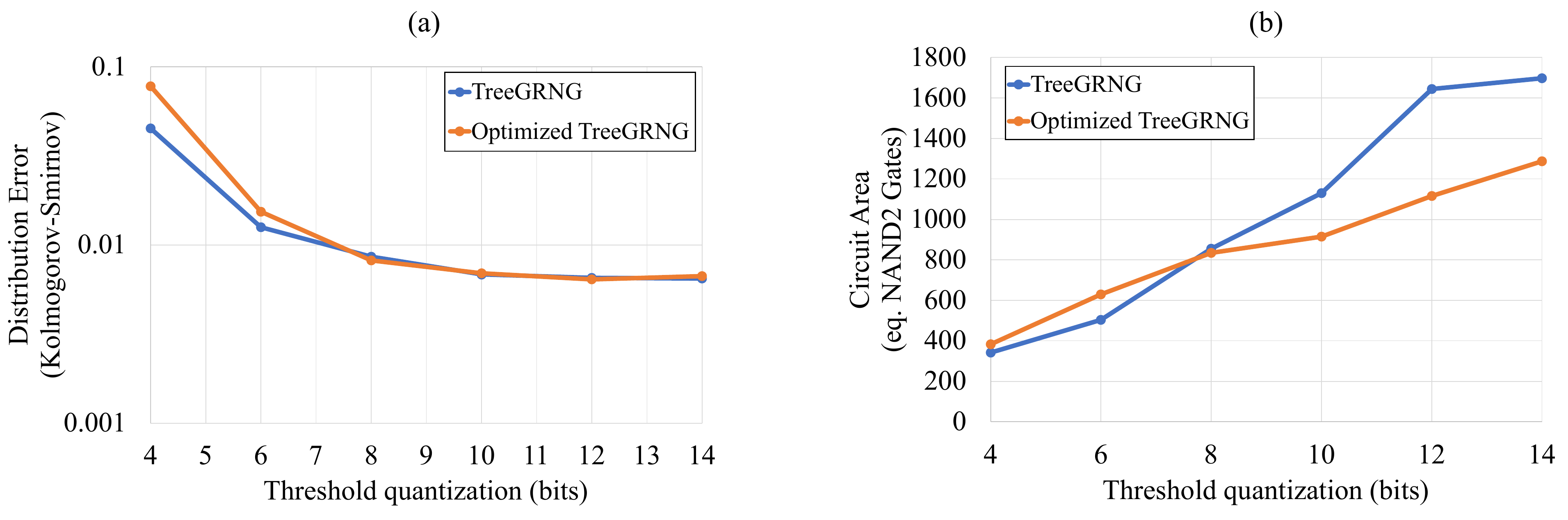}}
    \caption{\small{Circuit area (a) and error (b) versus the threshold quantization for our TreeGRNG proposal and its Optimized version.}}
    \label{clusterimprov}
    \vspace{-.5cm} 
\end{figure*}

The Kolmogorov-Smirnov Test (KST) is used to express the quality of the resulting distribution.
The KST corresponds to the biggest error made on the cumulative distribution function (CDF) curve when comparing the approximated distribution with the golden distribution \cite{b8}.
Fig.~\ref{clusterimprov}-a shows the results of this test when measuring the error difference between an 8-bit GRNG \mv{(8 levels, N=8)}, that uses the additional Hardware-reducing optimizations mentioned in Section~\ref{improv} with an implementation that does not use any optimizations. The thresholds that are used to represent the probabilities of the weighted coinflips are \mv{quantized with a} bitwidth ranging from 4 bits up to 14 bits. 
As both the predetermining of the thresholds and the use of the Gaussian symmetry are exact, both optimizations can be used without any drawbacks in the error performance. 
The clustering of the neighboring bin's threshold on the other hand simplifies the architecture by actually manipulating the value of the thresholds, which could lead to increased error.
Results show that this error \mv{is negligible for threshold quantization precisions of 8-bit and beyond, which is also the precision required to let the 8-level GRNG reach its best Gaussian generation quality}. 

\subsection{Hardware performance}

The GRNG hardware architecture proposal, along with the related work, has been implemented in SystemVerilog and synthesized using Cadence Genus. 
To assess circuit performance and cost, and to facilitate future work comparisons, we utilize the Silvaco\textsuperscript{\tiny{TM}} 45nm Open-source standard cell library. 
Additionally, we include the normalized circuit area, quantified in terms of its equivalence to NAND2$\_$X1 gates.
The hardware performance \mv{was assessed both} for implementations with various threshold lengths, as well as with and without the proposed TreeGRNG optimizations. 
While the proposed optimizations showed little influence on the error, Fig.~\ref{clusterimprov}-b demonstrates that significant improvements happen in the circuit area. 
\mv{The threshold clustering and the Gaussian symmetry drastically reduce the hardware cost in the region of interest beyond 8-bit thresholds, while minimally affecting accuracy in this region.} 
Also clear in Fig.~\ref{clusterimprov}-b is that the circuit area only grows linearly with the threshold size. 
This makes sense as only the comparators grow in size, proportionally to the grow in threshold width. 

Unlike the SoTA RNGs, our TreeGRNG algorithm offers the flexibility of generating random samples with any desired PDF distribution at the runtime by employing programmable thresholds.
However, such a programmable implementation requires \mv{threshold-programmable} comparators for every splitting probability, costing a \mv{significant additional} area.
Therefore, we can extract extra benefits for TreeGRNG implementation by hardwiring the thresholds of the desired PDF at the design time, since GRNG always generates Gaussian distributions.
Herein, we investigate the benefits in the circuit area of adopting hardwired thresholds and respectively the price of a programmable implementation.
Results show that the programmable TreeGRNG implementation uses an area equivalent to 3152 gates, while the hardwired constant thresholds reduce it to 1116 gates, an impressive 3$\times$ less hardware. 
Predetermining the thresholds at design time to use ultra-low-cost \mv{hardwired} comparators is key to achieving extremely highly efficient hardware results in the TreeGRNG architecture. 
It is worth noting that if a designer desires the flexibility to define the PDF distribution at runtime, our algorithm offers this groundbreaking solution, marking it, to the best of our knowledge, the first of its kind in the literature.



\begin{figure}[h]
    \centering
    \vspace{-.7cm}
    \includegraphics[width=3.3in]{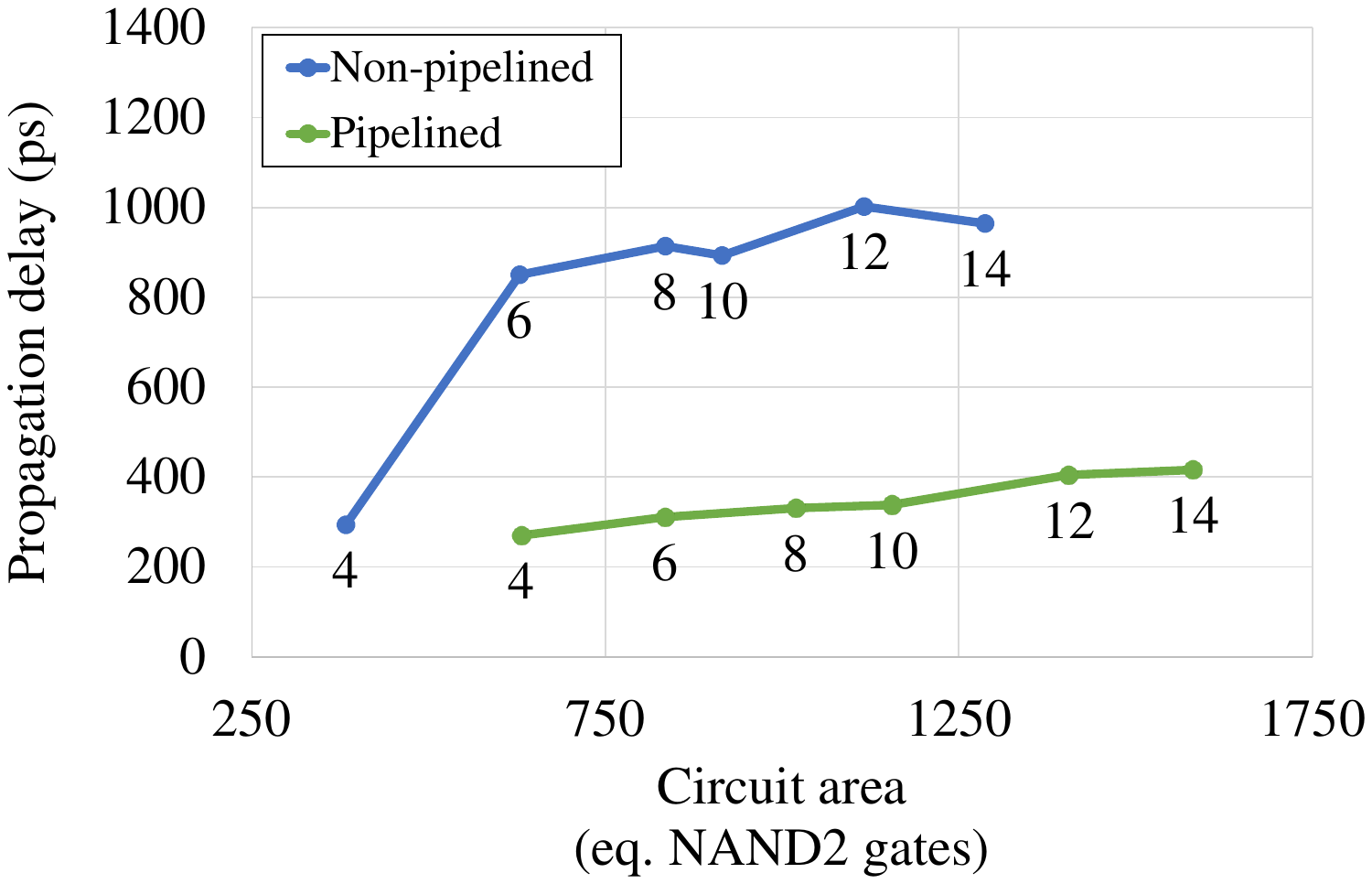}
    \vspace{-.7cm}
    \caption{\small{Propagation delay for the GRNG architectures.}}
    \label{delay}
    \vspace{-.05cm}
\end{figure}

As mentioned in Section~\ref{improv}, it is easily possible to pipeline the proposed algorithm
to shorten the critical path and achieve higher throughput. 
The \mv{8-level GRNG with different threshold quantization precisions} was pipelined and compared to its non-pipelined
counterparts. Given the fact that the amount of pipelining registers is \mv{mostly} independent of the \mv{threshold} bitwidth, 
the additional hardware cost is largely constant. An area equal to about 250 gates needs to be added to pipeline an 8-bit GRNG.
As seen in Fig.~\ref{delay}, pipelining improves performance by a factor of up to $2.5 \times$.
Considering both the additional hardware cost and timing improvements, pipelining seems a viable option when working with longer bitwidth thresholds where the additional hardware cost would be small compared to the circuit area of the complete system.

\begin{figure*}[t]
    \centering
    \vspace{-1.6cm}
    \includegraphics[width=.9\textwidth]{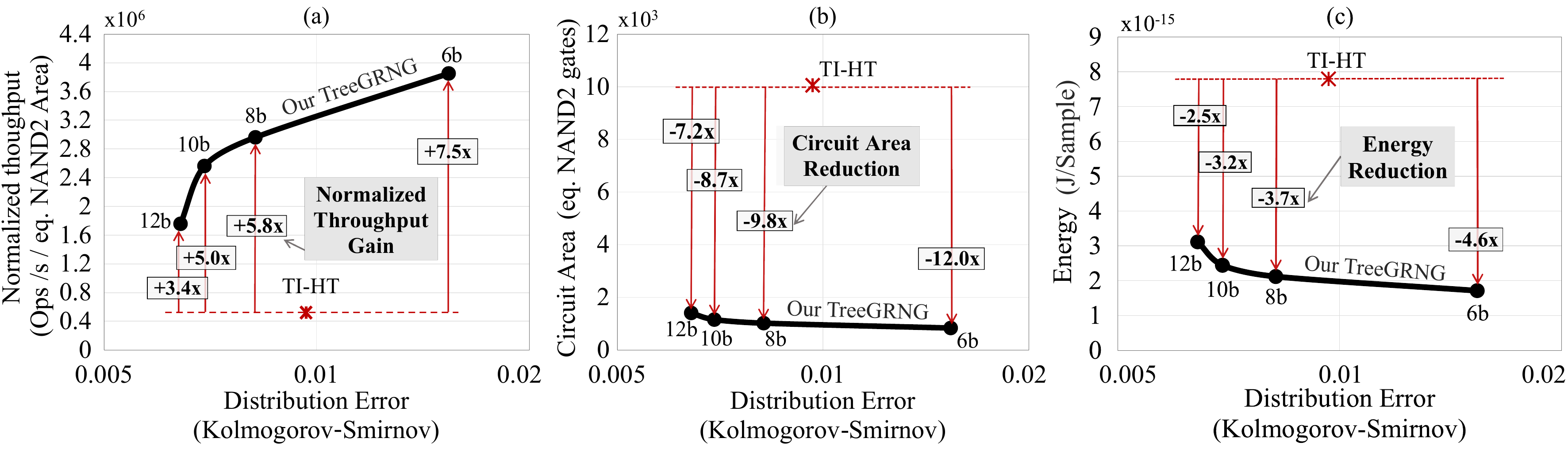}
    \caption{\small{Comparison between our pipelined TreeGRNG and the related work in TI-HT \cite{b6}.}}
    \label{speed}
    \vspace{-.3cm}
\end{figure*}

\begin{table*}[b]
    \vspace{-.4cm}
    \caption{\small{Comparison with the State of The Art.}}
    \vspace{-0.35cm}
    \begin{center}
    \begin{tabular}{c c c c c c c c}
    \hline
    \hline
    \textbf{Design*}& {\begin{tabular}{@{}c@{}}\textbf{Samples} \\\textbf{per Cycle} \end{tabular}} & {\begin{tabular}{@{}c@{}}\textbf{Latency} \\\textbf{(Cycles)} \end{tabular}} & {\begin{tabular}{@{}c@{}}\textbf{Energy} \\\textbf{(Joule/Sample)} \end{tabular}}&{\begin{tabular}{@{}c@{}}\textbf{Throughput} \\\textbf{(MSamp/sec)} \end{tabular}}&{\begin{tabular}{@{}c@{}}\textbf{Normalized throughput} \\\textbf{(kSamp/sec/NAND2-eq.)} \end{tabular}}&\addstackgap{\begin{tabular}{@{}c@{}}\textbf{Circuit Area} \\\textbf{(NAND2-eq.)} \end{tabular}}&\addstackgap{\begin{tabular}{@{}c@{}}\textbf{Error} \\\textbf{(KST)} \end{tabular}} \\
    \hline
    \rule{0pt}{4pt}
    \addstackgap[8pt]TI-HT\cite{b6}& 4&16 & $7.80*10^{-15}$&5168&514&10055&0.00966\\
    \Tgrline\addstackgap{\begin{tabular}{@{}c@{}}\textbf{Our TreeGRNG} \\ \end{tabular}}& 1 &1 &$2.48*10^{-15}$&1094&1310&835& \\
    \Tgrline\addstackgap{\begin{tabular}{@{}c@{}}\textbf{Our TreeGRNG} \\ \textbf{Pipelined} \end{tabular}} &1 &2 &$2.12*10^{-15}$&3021&2962&1020&\mr{-3}{*}{0.00820}\\
    \hline
    \hline
    \end{tabular}
    \label{stattable}
    \end{center}
    \vspace{-.1cm}
    \begin{tabular}{cccccccc}
    \mc{8}{l}{ \ \ \ \ \ \ \ \ \ \scriptsize *All designs, including the related work are synthesized in the same technology within an INT8 sample precision, employing the Silvaco\textsuperscript{\tiny{TM}} 45nm standard cells.}
    \end{tabular}
        \vspace{-1.2cm}
\end{table*}

\subsection{SotA comparison}

To demonstrate the pros and cons of \mv{the TreeGRNG, we compare it to} a replicated \mv{implementation} of the SoTA central limit theorem-based TI-HT \mv{GRNG} \cite{b6}. The TI-HT \mv{was selected as the most optimal SotA implementation for comparison, as it is the only other SotA solution that avoids multipliers and other complex functions (Table~\ref{tab:RelatedWorkResults}).} 
To keep the architecture comparison fair, the TI-HT algorithm from \cite{b6} was \mv{also implemented to generate} output samples with an INT8 precision, 
\mv{and using} the same standard cell \mv{library for synthesis}. 
We evaluate the results for both the non-pipelined and pipelined versions of the TreeGRNG algorithm. 
The required memory to implement TI-HT is not \mv{accounted for} in the total area. 

Fig.~\ref{speed} shows the comparison results between the TI-HT and the pipelined TreeGRNG with all optimizations. The TreeGRNG is represented by different threshold quantizations. Results are shown for 6, 8, 10, and 12-bit thresholds. Fig.~\ref{speed}-a shows the gain in normalized throughput, \mv{demonstrating a} 
3.4$\times$ \mv{improvement for the fully accurate} 12-bit~threshold, all the way to a 7.5$\times$ improvement for the 6-bit~thresholds. Similarly, Fig.~\ref{speed}-b shows the area results, with \mv{improvements ranging from} 7.2$\times$ to an impressive 12$\times$. Lastly, Fig.~\ref{speed}-c shows the important metric of Energy per Sample. Analogous to both earlier comparisons, the TreeGRNG requires up to 4.6$\times$ less energy for 6-bit thresholds and still requires 2.5$\times$ less energy for the 12-bit threshold. 
\mv{It is clear that tuning the threshold precision allows to trade-off between} high throughput lower accuracy samples, and lower throughput high accuracy samples.

To compare with the SoTA at a similar error level, we select the version with the 8-bit threshold quantization, based on the results from Fig.~3-a). Table~\ref{stattable} shows the results for energy, throughput, throughput per unit area, and the circuit area for both the pipelined and non-pipelined architecture.
\mv{At 8-bit thresholds,} the TreeGRNG reaches 3.7$\times$ higher energy efficiency than the TI-HT.
In contrast to TI-HT, our advantage is largely due to the simplicity of the algorithm where no arithmetic units are required. 
At the cost of 10$\times$ larger circuit area, an INT8 TI-HT implementation only reaches a throughput less than 2$\times$ higher than our pipelined TreeGRNG proposal. 
Therefore, our pipelined TreeGRNG reaches a lower error and generates samples at a 5.8$\times$ higher normalized throughput per unit area than the TI-HT.

\section{Conclusion}
This paper introduced the TreeGRNG, an approach for efficiently generating \mv{gaussian random numbers}. 
Our algorithm is inspired by a binary tree where each level generates a bit by performing a weighted coinflip, while each probability is determined by the result of previous coinflips. 
The GRNG architecture is efficiently implemented by calculating the \mv{required} probabilities 
at design time, allowing the synthesize tool to reduce the comparators' circuit area footprint to a minimum.
Our design distinguishes itself from SoTA as it does not require any complex function or arithmetic units to get accurate results. 
This avoids the use of LUTs or complex arithmetic, making our architecture more area- and power-efficient.
Multiple implementations were tested with different threshold quantization and compared to SoTA. 
While this paper focuses on GRNG generation, the approach can also be used to generate any PDF distribution. 

\section*{Acknowledgment}
This project has been partly funded by the European Research Council (ERC) under grant agreement No. 101088865, the European Union’s Horizon 2020 program under grant agreement No. 101070374, the Flanders AI Research Program, and KU Leuven. Guilherme Paim expresses gratitude to the CAPES Brazilian Foundation for their financial support in past grants and for accepting the novation agreement that allowed him to work abroad.

\end{document}